\documentclass{article}

\usepackage{microtype}
\usepackage{graphicx}
\usepackage{subfigure}
\usepackage{booktabs} 
\usepackage{amsmath}
\usepackage{amsthm}
\usepackage{amsfonts}
\usepackage{optidef}
\usepackage{siunitx}

\usepackage{color}
\usepackage{regexpatch} 
\usepackage[colorinlistoftodos,prependcaption]{todonotes}
\makeatletter
\xpatchcmd{\@todo}{\setkeys{todonotes}{#1}}{\setkeys{todonotes}{inline,#1}}{}{}
\makeatother

\usepackage{amsmath,amsfonts,bm}

\def\eqref#1{equation~\ref{#1}}

\def\1{\bm{1}}

\DeclareMathAlphabet{\mathsfit}{\encodingdefault}{\sfdefault}{m}{sl}
\SetMathAlphabet{\mathsfit}{bold}{\encodingdefault}{\sfdefault}{bx}{n}

\newcommand\norm[1]{\left\lVert#1\right\rVert_{L^2}}

\def\J{{\mathcal{J}}}
\def\La{{\mathcal{L}}}
\def\H{{\mathcal{H}}}

\newtheorem{theorem}{Theorem}

\usepackage{hyperref}

\usepackage{caption}

\usepackage[accepted]{icml2019}

\icmltitlerunning{Learning Dynamical Systems from Partial Observations}

\begin{document}

\twocolumn[

\icmltitle{Learning Dynamical Systems from Partial Observations}


\icmlsetsymbol{equal}{*}

\begin{icmlauthorlist}
\icmlauthor{Ibrahim Ayed}{equal,lip6,thales}
\icmlauthor{Emmanuel de B\'ezenac}{equal,lip6}
\icmlauthor{Arthur Pajot}{lip6}
\icmlauthor{Julien Brajard}{bergen,locean}
\icmlauthor{Patrick Gallinari}{lip6,criteo}
\end{icmlauthorlist}

\icmlaffiliation{lip6}{Sorbonne Universit\'e, UMR 7606, LIP6, F-75005 Paris, France}
\icmlaffiliation{thales}{Theresis lab, Thales, Thales Research \& Technology Route D\'epartementale, 91120 Palaiseau}
\icmlaffiliation{locean}{Sorbonne Universit\'e, CNRS-IRD-MNHN, LOCEAN, Paris, France}
\icmlaffiliation{bergen}{Nansen Environmental and Remote Sensing Center, Bergen, Norway}
\icmlaffiliation{criteo}{Criteo AI Lab, Paris, France}
\icmlcorrespondingauthor{Ibrahim Ayed}{ibrahim.ayed@lip6.fr}
\icmlcorrespondingauthor{Emmanuel de B\'ezenac}{emmanuel.de-bezenac@lip6.fr}

\icmlkeywords{Machine Learning, ICML}

\vskip 0.3in
]



\printAffiliationsAndNotice{\icmlEqualContribution} 

\begin{abstract}

We consider the problem of forecasting complex, nonlinear space-time processes when observations provide only partial information of on the system's state. We propose a natural data-driven framework, where the system's dynamics are modelled by  an unknown time-varying differential equation, and the evolution term is estimated from the data, using a neural network. Any future state can then be computed by placing the associated differential equation in an ODE solver. We first evaluate our approach on shallow water and Euler simulations. We find that our method not only demonstrates high quality long-term forecasts, but also learns to produce hidden states closely resembling the true states of the system, without direct supervision on the latter. Additional experiments conducted on challenging, state of the art ocean simulations further validate our findings, while exhibiting notable improvements over classical baselines.\\

\end{abstract}

\section{Introduction}

Dynamical systems are a tool of choice to model the evolution of phenomena occurring in nature. In order to derive a dynamical system describing a real world physical process, one must first gather measurements of this system. Then, a set of variables $X_t $ describing the system at a given time $t$, called the \textit{state}, along with a transition function $T(X_t) = X_{t+\delta t}$ linking consecutive states in time, is inferred based on the available measurements. Generally, the continuous limit proves to be more tractable, powerful and convenient for calculations, so that one usually considers an evolution equation of the form~:
\begin{equation}
\label{eq:state}
    \frac{dX_t}{dt} = F(X_t)
\end{equation}
Many phenomena studied in physics, computer vision, biology~\cite{pde_biology}, geoscience~\cite{pde_geoscience}, finance~\cite{pde_finance}, etc... obey a general equation of this form. For this reason, an extensive effort has been put into gaining a better understanding and resolving this equation. However, for many practical problems, the relation between the components of the state is highly non-linear and complex to describe analytically: finding an appropriate evolution model $F$ can thus elude scientific communities for decades.

With the availability of very large amounts of data captured via diverse sensors and recent advances of statistical methods, a new data-driven paradigm for modeling dynamical systems is emerging, where relations between the states are no longer handcrafted, but automatically discovered based on the available observations. This problem can be approached by considering some class of admissible functions $\{F_\theta\}$, and looking for a $\theta$ such that the solution $X^\theta$ of~:
\begin{equation}
\label{eq:state_nn}
    \frac{dX_t}{dt} = F_\theta(X_t)
\end{equation}
fits the measured data. This approach has motivated some recent work for exploiting machine learning in order to solve differential equations. For example, \citet{rudy_data-driven_2017} parameterizes $F_\theta$ as sparse linear regression over a set of pre-defined candidate differential terms, \citet{RAISSI2017683,Raissi18} or \citet{Long2018} use statistical models such as Gaussian processes and neural networks to model $F_\theta$ and learn a solution to the corresponding equation.

Previous methods have essentially considered the case where the state of the system $X_t$ is fully-observed at all times $t$. However, for many real-world applications, the entire state of the system is not fully visible to external sensors: one usually only has access to low-dimensional projections of the state, \textit{i.e.} \textit{observations}. Intuitively, the latter can be seen as what is readily and easily measurable; this means that, in contrast with the ideal case where the full state can be observed at all times with perfect certainty, there is an important loss of information. This issue is a major one in many fields within applied sciences \cite{data_assim_bocquet, CAnalysisMF}.

In our work, we consider the problem of learning complex spatio-temporal dynamical systems with neural networks from observations $Y$, which are only partially informative with respect to the full state $X$. First, we formulate the problem as a continuous-time optimal control problem, where the parameters of the neural network are viewed as control variables. From this, we then present a natural algorithm solving the resulting optimization problem, placing the neural network in an ordinary differential equation (ODE) solver in order to produce future predictions. Finally, we successfully apply our method to three increasingly challenging datasets and show promising results, comparing our approach to standard deep learning baselines.
 
Our main contributions are the following:
\vspace{-0.3cm}
\begin{itemize}
\itemsep0em 
    \item[--] a general, widely applicable approach for modeling space-time evolving processes with neural networks;
    \item[--] linking the continuous-time optimal control framework to neural network training in the partially observed case;
    \item[--] experiments with realistic dynamical systems exhibiting good forecasting performance for long time horizons in different settings;
    \vspace{-0.1cm}
    \item[--] experiments showing successful unsupervised learning of the true hidden state dynamics of the dynamical system;
    \vspace{-0.1cm}
    \item[--] all our results are achieved without imposing priors over the form of the studied equation. This is an important novelty w.r.t. existing work. 
\end{itemize}
\vspace{-0.2cm}

\section{Background}

\subsection{Continuous State Space Models}

We consider space-time dynamics for which $X$ can be written as a function of $(t,x) \in \mathbb{R}_+ \times \Omega$ where $t$ and $x$ are respectively the time and space variables, $\Omega\subset\mathbb{R}^d$ the domain over which we study the system. The spatial vector-valued function $X_t$ contains the quantities of interest describing a studied physical system at time $t$. 

In a realistic setting, the state is generally only partially observed \textit{e.g.}, when studying the ocean's circulation, variables contained in the system's state such as temperature or salinity are observable, while others such as velocity or pressure are not. In other words, the measured data is only a projection of the complete state $X_t$. This measurement process can be modelled with a fixed operator $\H$ linking the system's state $X_t$ to the corresponding observation $Y_t$:
\[
Y_t=\H(X_t)
\]
In the following, $\H$ is supposed known, fixed and differentiable\footnote{In most practical cases, this hypothesis is verified as $\H$ can usually be represented as a smooth operator.}. Let us note that, generally, the measurement process represents a considerable loss of information compared to the case where $X$ is available, as the measurements may be sparse and low-dimensional.

Moreover, we assume that $X$ obeys a differential equation of the general form of \eqref{eq:state}, with an initial condition $X_0$. This leads us to the following continuous state space model:
\begin{equation}
\label{eq:statessystem}
  \left\{
  \begin{aligned}
    X_0\\ 
    \frac{dX_t}{dt} &= F(X_t)\\
    Y_t &= \H(X_t)\\
 \end{aligned}
\right.
\end{equation}

\subsection{Neural Ordinary Differential Equations}

Recently, the link has been made between residual networks and dynamical systems \citet{E2017}: a residual block ${h_{t+1} = h_t + f(h_t, \theta_t)}$ can be seen as the explicit Euler discretization of the following system:
\begin{equation}
    \label{eq:node}
    \frac{dh_t}{dt} = f(h_t, \theta_t)
\end{equation}
Adopting this viewpoint, time $t$ corresponds to the neural network's layer index, the initial condition $h(0)$ to the network's input, and the forward pass as the time integration ${h(T) = h(0) + \int_0^T f(h(t), \theta_t)\, \mathrm{dt}}$, where $h(T)$ corresponds to its output. \citet{node} propose computing this intractable integral using an ordinary differential equation (ODE) solver. During training, in order to compute the derivative with respect to the neural network parameters, the corresponding adjoint state equation is solved backward in time. Note that in this work, instead of considering the evolution of the inner dynamics of the neural throughout its layers, we consider the dynamics of the studied process itself, in the context partially observed states.

\section{Theoretical Framework}
\label{framework}

In this section, we set the theoretical framework necessary to solve our problem. As we will see, it can be formulated as a continuous-time optimal control problem, where the control variables correspond to the network's parameters. In order to train our model, we derive the forward and backward equations necessary for the classical gradient descent algorithm solving it and discuss the two main methods available to compute numerical solutions.

\subsection{Optimization Problem}

Our goal is to learn the differential equation driving the dynamics of a smooth state function $X$ for which we only have supervision over observations $Y$ through a fixed operator $\mathcal{H}$. In order to enforce our dynamical system to explain the observations, we define a cost functional of the form~:
\begin{equation}
\begin{split}
\J(Y,\widetilde{Y}) = \int_0^T \!  \norm{Y_t - \widetilde{Y}_t}^2 \mathrm{dt}
\label{eq:j}
\end{split}
\end{equation}
Here, $Y$ is a spatio-temporal field representing  observations of the underlying system, $\widetilde{Y}$ the output of the system, and $\norm{\cdot}$ the norm associated to the $L^2$ Hilbert space over $\Omega$.

Since the state $X_t$ is constrained to follow the dynamics described by equation \ref{eq:state_nn}, starting from its initial condition $X_0$, the optimization problem is in fact a constrained one~:
\begin{mini}|l|
  {\theta}{\mathbb{E}_{Y\in\text{Dataset}}\left[\J(Y,\H(X))\right]}{}{}
  \addConstraint{\dfrac{dX_t}{dt}}{=F_\theta(X_t)}{}
  \addConstraint{X_0}{=g_\theta(Y_{-k},\breve{X}_0)}{}
  \label{eq:optim_pb}
\end{mini}
where $F_\theta$ is a smooth vector valued function defining the trajectory of $X$, and $g_\theta$ gives us the initial condition $X_0$. In other words, $\theta$ parameterizes both the dynamics through $F$ and the initialization through $g$. In particular, if a full initial state is given as input to the system, $g_\theta$ can be taken as independent of any parameter and doesn't need to be learned.

For any $\theta$, we assume that $F$ and $g$ are such that there always exists a unique solution to the equation given as a constraint in \eqref{eq:optim_pb}. In the following, we will call such a solution $X^\theta$.

\subsection{Adjoint State Method}

Now that the optimization problem is stated, an algorithm to solve it must be designed. For this, we will use a standard gradient descent technique. In order to use gradient descent, we must first calculate the gradient of the cost functional under the constraints, \textit{i.e.} the differential of $\theta\to\mathbb{E}_Y\J(Y,\H(X^\theta))$. However, this implies calculating $\dfrac{\partial X^\theta}{\partial \theta}$, which is often very computationally demanding, as it implies solving $\dim(\theta)$ forward equations.

But, by considering the Lagrangian formulation of the constrained optimization problem introduced in \eqref{eq:optim_pb}, it is possible to avoid explicitly calculating  $\dfrac{\partial X^\theta}{\partial \theta}$. The Lagrangian is defined as~:
\begin{equation}
\begin{split}
\La(X, \lambda, \mu, \theta) = \;& \J(X) + \int_0^T  \left \langle \lambda_t, \frac{dX_t}{dt} - F_\theta(X_t) \right \rangle \, \mathrm{dt}\\
&+ \left \langle  \mu, X_0 - g_\theta \right \rangle
\label{eq:lagrangian}
\end{split}
\end{equation}
here, the scalar product $\left \langle \cdot, \cdot \right \rangle$ is the scalar product associated to the $L^2$ space over $\Omega$.

As, for any $\theta$, $X^\theta$ satisfies the constraints by definition, we can now write~:
\[
\forall \theta,\lambda, \mu,\ \La(X^\theta,\lambda,\mu,\theta) = \J(X^\theta)
\]
which gives~:
\[
\forall \lambda, \mu,\ \dfrac{\partial}{\partial \theta}\La(X^\theta,\lambda,\mu,\theta) = \dfrac{\partial}{\partial \theta}\J(X^\theta)
\]

By calculating the differential of $\La$ \textit{w.r.t.} $\theta$ and using it to have the gradient of $\J$, we can obtain~:

\begin{theorem}[Adjoint State Equation]
\label{th:theorem1}
    \begin{equation}
    \label{eq:grad}
        \dfrac{\partial}{\partial \theta}\J(X^\theta) = -\int_0^T \left \langle  \lambda_t, \partial_\theta F_\theta(X_t^\theta) \right \rangle\mathrm{dt} - \left \langle \lambda_0, \partial_\theta g_\theta \right \rangle
    \end{equation}
    where $\lambda$ is solution of~:
    \vspace{-0.1cm}
    \begin{equation}
    \label{eq:adjoint}
        \partial_t \lambda_t = A_t \lambda_t + B_t
    \end{equation}
    solved backwards, starting with $\lambda_T = 0$, and where~:
    \vspace{-0.2cm}
    \[
    A_t = -(\partial_X F_\theta(X_t^\theta))^\star
    \]
    \vspace{-0.15cm}
    and
    \vspace{-0.1cm}
    \[
    B_t = 2(\partial_X\H(X_t^\theta))^\star(\H(X_t^\theta)-Y_t)
    \]
    where $M^\star$ denotes the adjoint operator of linear operator $M$.
\end{theorem}

\vspace{-0.35cm}
\paragraph{Proof.} The proof is deferred to section of the supplementary material, Section \ref{proof}.

We now have equations entirely characterizing the gradient of our cost functional: for a given value of $\theta$, we can solve the forward \eqref{eq:state_nn} to find $X^\theta$. Then, $\lambda$ can be solved backwards as its equation only depends on $X^\theta$ which gives us all necessary elements to calculate the gradient of $\J$. This gives us the following iterative algorithm to solve the optimization problem, starting from a random initialization of $\theta$~:
\vspace{-0.1cm}
\begin{enumerate}
    \setlength\itemsep{0cm}
    \item Solve the forward state equation \eqref{eq:state_nn} to find $X^\theta$~;
    \item Solve the backward adjoint equation \eqref{eq:adjoint} to find the corresponding $\lambda$~;
    \item Update $\theta$ in the steepest descent direction using equation \eqref{eq:grad}.
\end{enumerate}

From these steps (and taking into account the estimation of the initial state, further explained in Section \ref{sec:expes}), we can derive an algorithm for training, summarized in Algorithm~\ref{algo}.

\begin{algorithm}
  \caption{Training Procedure}
\begin{algorithmic}
\label{algo}
  \STATE {\bfseries Input:} Training samples $\{(Y_{-k}, \breve{X_0}, ), Y_{+l}\}$.
  \STATE Guess initial parameters $\theta$
  \WHILE{not converged}
  \STATE Randomly select sample sequence $\{(Y_{-k}, \breve{X_0}, ), Y_{+l}\}$
  \IF{Initial State is Fully Observed}
    \STATE $X_0 \gets \breve{X_0}$
  \ELSE
    \STATE $X_0 \gets g_\theta(Y_{-k}, \breve{X_0})$ \hspace{0.15cm} 
  \ENDIF
    \STATE Solve Forward $\frac{dX_t}{dt} = F_\theta(X_t), \; 
    X(0) = X_0, \; t \in [0, l]$
  \STATE Solve Backward $\dfrac{d\lambda_t}{dt} = A_t \lambda_t + B_t, \; \lambda_l=0, \; t \in [0, l]$
  \STATE Compute gradient $\frac{\partial \mathcal{J}}{\partial \theta}(X^\theta)$
  \STATE Update $\theta$ in the steepest descent direction
  \ENDWHILE
\STATE {\bfseries Output:} Learned parameters $\theta$.
\end{algorithmic}
\end{algorithm}
\vspace{-0.1cm}

\subsection{Approximate Solutions}

While Algorithm \ref{algo} seems quite straightforward, solving the forward and backward equations (\ref{eq:state_nn}, \ref{eq:adjoint}) generally is not. Typically, they do not yield a closed form solution. We must content ourselves with approximate solutions. There are essentially two different ways to tackle this problem \cite{control}: the \textit{differentiate-then-discretize} approach, or the \textit{discretize-then-differentiate} approach\footnote{The differentiate-then-discretize method is often referred to as the \textit{continuous adjoint method}, and the \textit{discretize-then-differentiate} approach as the \textit{discrete adjoint method}~\cite{sirkes1997finite}.}.

In a \textit{differentiate-then-discretize} approach, one directly approximates the equations using numerical schemes. Here, the approximation error to the gradient comes from the discretization error made in the solver for both the forward and backward equations. This method is used in the black box solvers presented in \citet{node}. This method has the advantage of allowing the use of non-differentiable steps in the solver. However, this method can yield inconsistent gradients of cost functional $\J$, and the discretization of the adjoint equations depends highly on the studied problem and must carefully be selected~\cite{Bocquet2012}.

In a \textit{discretize-then-differentiate} approach, a differentiable solver for the forward equations is used, \textit{e.g.} using an explicit Euler scheme $X^\theta_{t+\delta t} \approx X^\theta_t + \delta t F_\theta(X^\theta_t)$. Based on the solver's sequence of operations for the forward equations, the backward equations and the gradient can be directly obtained using automatic differentiation software \cite{pytorch}. This algorithm is actually equivalent to backpropagation \cite{lecun1988theoretical}. As the stepsize approaches zero,  the forward and backward equations are recovered. In this paper, we will use this method as the explicit Euler solver gives good results for our examples while being more easily tractable.

\section{Experiments}
\label{sec:expes}
In this section we evaluate our approach, both quantitatively and qualitatively. We consider three different datasets representing dynamical systems with increasing complexity. We evaluate our method with respect to its ability to predict observations and to reproduce the dynamics of the hidden state. For the first two datasets, we use the full initial condition as input. For the last dataset, we only have access to a subset of the states which makes us propose a variant of our approach in order to accommodate this situation.

\subsection{Datasets}

The first two datasets are completely simulated: we have the true full state to initialize our algorithm $X_0$ in equation (\ref{eq:optim_pb}). The last dataset is based on a complex simulation, where real observations are assimilated in order to correct the simulation. Note that for this dataset, we do not have access to the full initial conditions.

\begin{itemize}
    \item \textbf{The Shallow Water equations} are derived from the Navier Stokes equations when integrating over the depth of the fluid (see supplementary material, section \ref{shalw}). These equations are discretized on a spatial $80 \times 80$ grid. We decompose the simulation into train-validation and test subsets of $600$ and $1000$ acquisitions respectively.
    \item \textbf{The Euler equations}, which are also derived from the Navier Stokes equations when neglecting the viscosity term (see supplementary material Section \ref{section:euler}). These equations are discretized on a spatial $64 \times 64$ grid. We use $15 000$ observations for the train set and $10 000$ for the test.

\item \textbf{Glorys2v4}, \citet{parent2013global} is a very challenging simulation to learn. We consider as observations the Sea Surface Temperature (SST) from a certain zone provided by the Global Ocean Physics reanalysis Glorys2v4 provided by the Copernicus Marine environment monitoring service~\footnote{\url{http://marine.copernicus.eu}}.  A brief description of Glorys2v4 is provided in appendix \ref{section:nemo}. The dataset consists of daily temperatures from 2006-12-28 to 2015-12-30, from which we extracted a $64 \times 64$ sub-region. We take the first $3000$ days for training, and leave the rest for the test set. Here, the full state is not completely available as initial input, we only have a proxy for one variable and for two dimensions of it: the velocity field. This makes initializing our dynamical system more challenging.

\end{itemize}

\subsection{Implementation Details}

We decompose the simulations into training sequences of fixed length, using $6$ timesteps for the target sequence. In practice, the cost functional $\J$ is estimated on a minibatch of sequences from the dataset, and optimized using stochastic gradient descent. 

Throughout all the experiments, $F_\theta$ is a standard residual network \cite{he_deep_2016}, with $2$ downsampling layers,  $6$ residual blocks, and bilinear up-convolutions instead of transposed convolutions. To discretize the forward equation (\ref{eq:state_nn}) in time, we use a simple Euler scheme. Note that the discretization stepsize may differ from the time interval between consecutive observations; in our case, we apply $3$ Euler steps between two observations, \textit{i.e.} $\delta t=1 / 3$. For the spatial discretization, we use the standard gridlike discretization induced by the dataset.

The weights of the residual network $\theta$ are initialized using an orthogonal initialization. Our model is trained using a exponential scheduled sampling scheme with exponential decay, using the Adam optimizer, with a learning rate set to $1\times10^{-5}$. We use the Pytorch deep learning library \cite{pytorch}.

\subsection{Experiments with Shallow water equations}
\begin{figure*}[ht]
\begin{center}
 \includegraphics[width=0.9\textwidth]{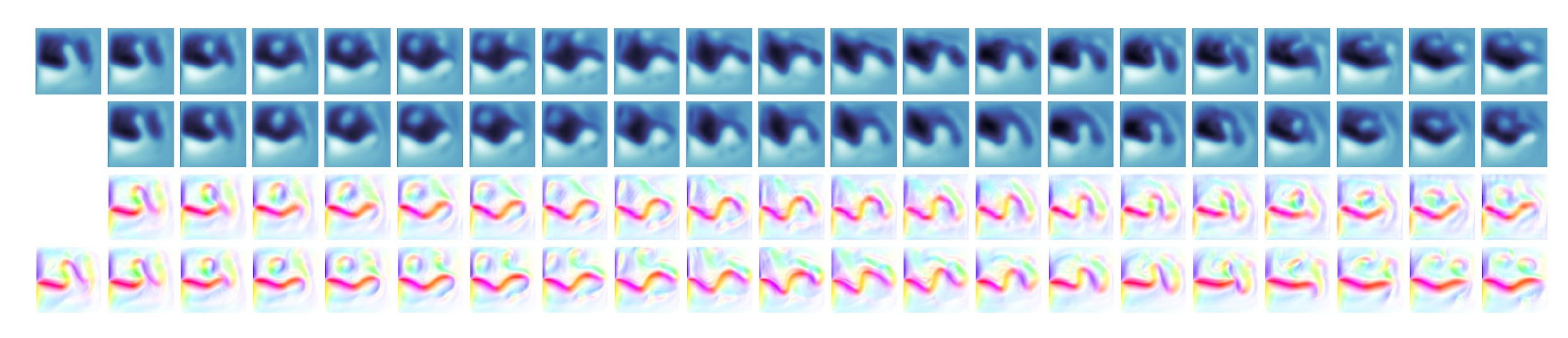}
\end{center}
\vspace{-0.5cm}
\caption{\label{Fig-SW}Forecasting the shallow water model on the test set. From top to bottom: input (top leftmost frame) and target observations, model output, model hidden state, and the two dimensional velocity vector, as input (left), and ground truth (right). By learning to forecast the observations, our model produces hidden states closely resembling the system's true hidden state, without direct supervision.}
\end{figure*}
The system of equations is described in more details in the supplementary material, Section \ref{shalw}. Here, the state $X$ corresponds to the column height and the two-dimensional velocity vector field, $\H$ is a linear projector giving the first component of $X$ so that observation $Y$ is the mixed layer depth anomaly and velocity is unobserved. The problem amounts to predicting future states with a training supervision over densities only and an initial full state $X_0$ given to the system. For experiments with shallow water and Euler simulations, we set $g_\theta=X_0$ to be equal to the initial full state provided as input to the system. Note that it is not uncommon to have prior knowledge on the system's initial condition state \cite{bereziat}.

\begin{figure*}[ht]
\begin{center}
 \includegraphics[width=0.9\textwidth]{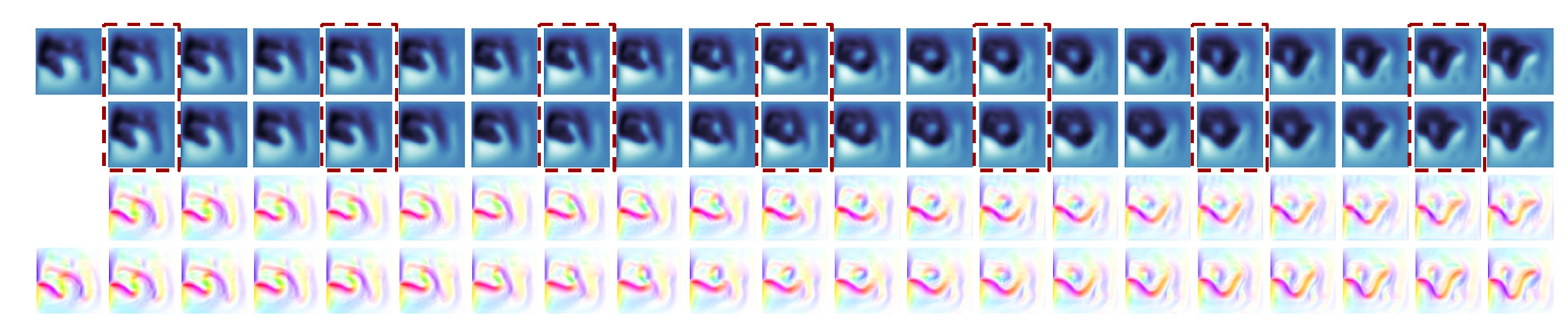}
\end{center}
\vspace{-0.5cm}
\caption{\label{Fig_time_int}Time interpolations with our approach on the test set. We train our model by regressing to the targets every 3 images (materialized by the red boxes). We then compare the outputs of the model with the unseen ground truth states. }
\end{figure*}

\paragraph{Forecasting Observations.} Figure \ref{Fig-SW} shows a sample of the predictions of our system over the test set. We can clearly see that it is able to predict observations up to a long forecasting horizon, which means that it generalizes and thus has managed to learn the dynamical system. Note that the initial state used at test time has never been seen at training time which means that the optimization problem was solved correctly without over-fitting. The cost function and the supervision were only defined at the level of observations. For the velocity vector field, color represents the angle, and the intensity the magnitude of the associated vectors.
\vspace{-.15cm}
\paragraph{Hidden State Discovery.}
Our method forecasts a full state $X_t$ and not only the observations $Y_t$. In order to predict the observations correctly, our model has to learn to predict future hidden states that contain information of the true state. By feeding the true initial conditions to our model, we find that our method is able to learn the true dynamics of the hidden state with a good accuracy, while never directly enforcing a penalty on the the latter. Note that the only access our method has to full states is through the initial state provided as input. This result is intriguing: the model should theoretically be able to use a state \textit{encoding} that is different from the one given by the initial condition. We hypothesize that our network's architecture is biased towards preservation of the input code. This is also empirically observed in the domain translation domain.

\paragraph{Interpolation between data points.}
Our framework allows us to forecast for arbitrary times $t$. Figure \ref{Fig_time_int} shows a sample of this interpolation mechanism. In this example, the model has been trained by regressing to the targets every 3 images (materialized on the figure by the red boxes). The outputs of the model are then compared with the unseen ground truth states. This shows that our approach allows us to learn the true evolution of the state. This is an important feature of our method, similar in this aspect to the claims of \citet{node}, even though it is applied here to a high-dimensional, highly non-linear and partially observed learned dynamical system, for which we can interpolate the observations as well as the inferred hidden state.

\subsection{Experiments with the Euler equations}

\begin{figure}[ht]
\begin{center}

 \includegraphics[width=0.45\textwidth]{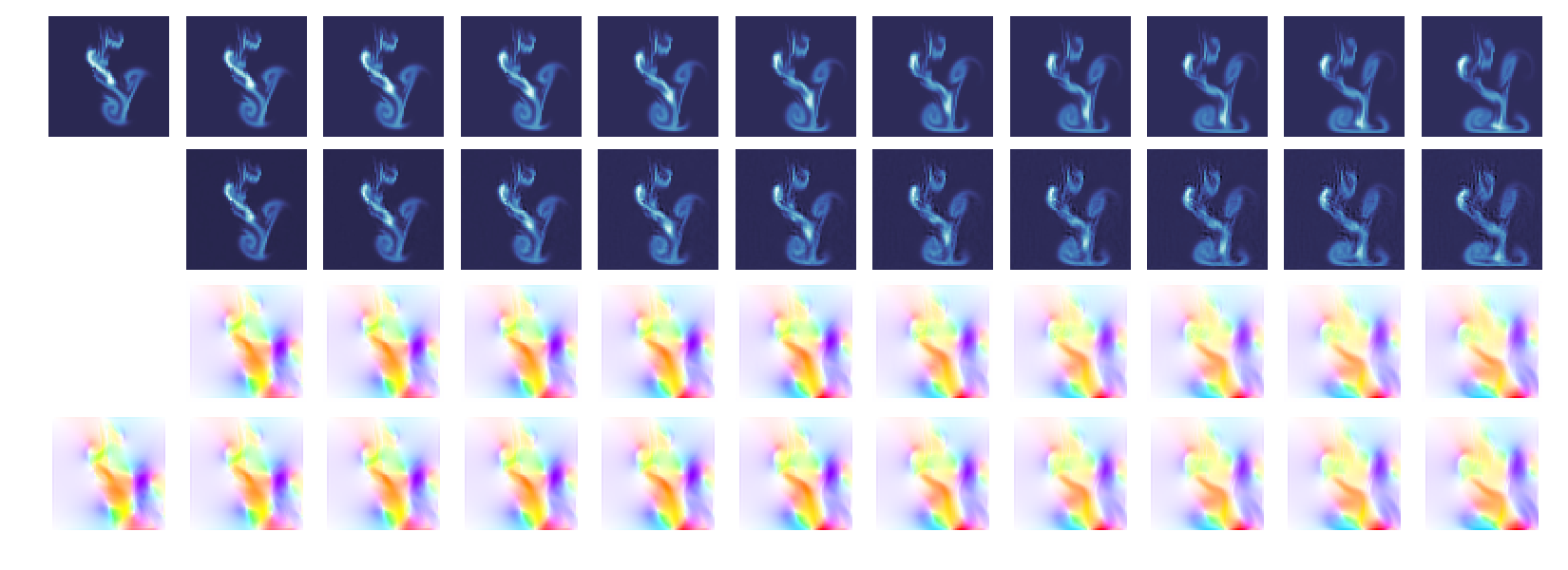}
\end{center}
\vspace{-0.5cm}
\caption{Forecasting the Euler equations on the test set. From top to bottom: input and target observations, model output, model hidden state, hidden state input and ground truth.}
\label{Fig_Euler}
\end{figure}

The encouraging results of the previous subsection made us want to try our methods with more complex dynamics, namely the Euler equations, in the same conditions to see if it is able to cope with a more difficult example. We use exactly the same architecture as the the previous experiment, and obtain similarly good results on the test set, as shown in Figure \ref{Fig_Euler}. Again, we manage to predict observations as well as hidden state dynamics for long forecasting horizons with one full state as input and a training supervision over observations only. The form of Euler equations is provided in appendix \ref{section:euler}.

\subsection{Experiments with Glorys2v4}

This dataset is much more challenging and represents a leap from the fully simulated ones presented before. One reason is obviously the high dimensionality of the system and the absence of a full state as initial input to our system as we only have a proxy over the velocity field. A second one is the fact that we only work over sequences from the same ocean zone while the model functions within a larger area. This makes the dynamics for a single zone non-stationary as boundary conditions are constantly shifting, thus violating an important assumption of our method and making it almost impossible to make long term forecasts with a reasonable number of observations. All we can hope for is for the dynamics to be locally stationary so that the model can work well for a few steps.

\begin{figure*}[ht]
\begin{center}
 \includegraphics[width=0.9\textwidth]{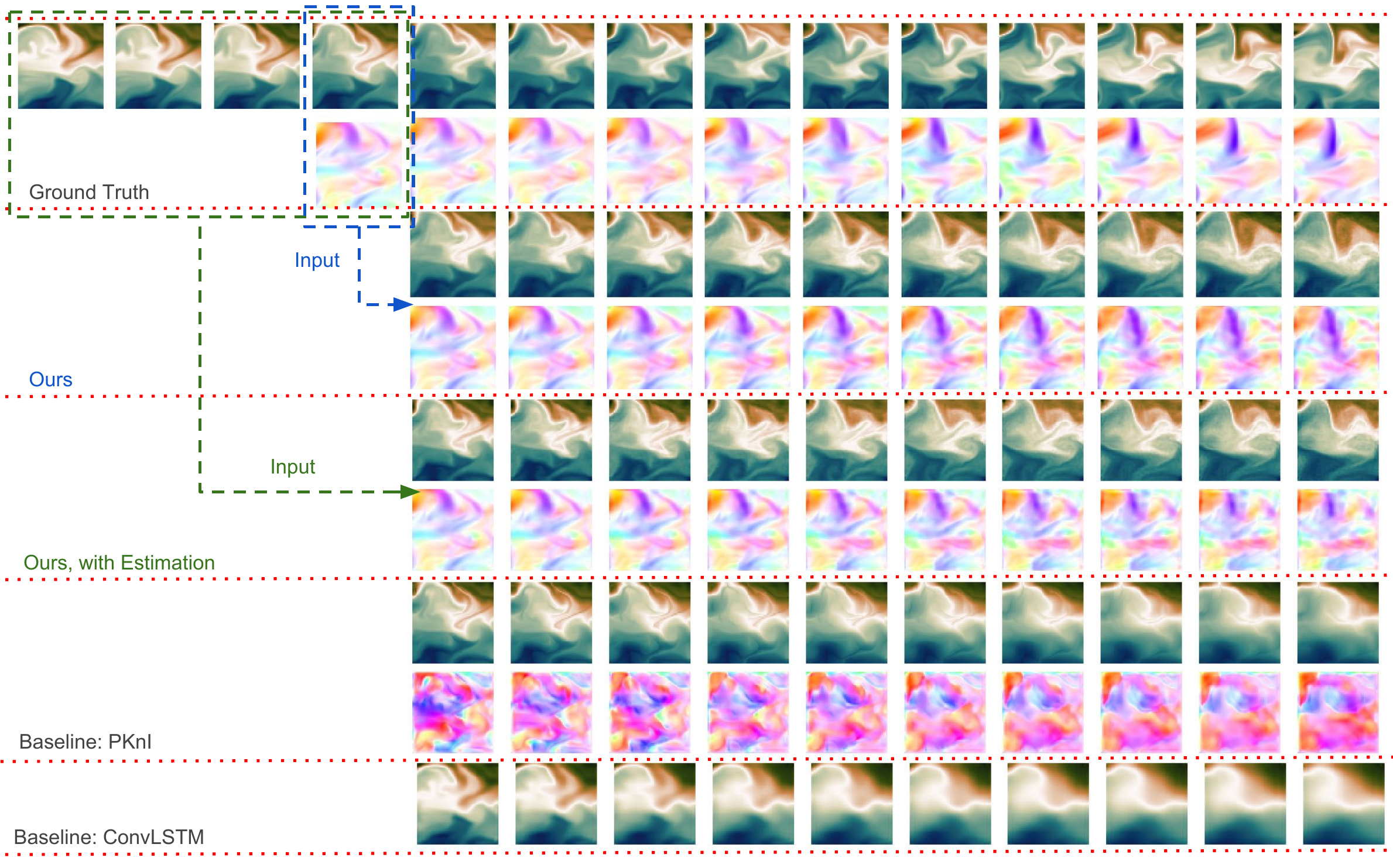}
\end{center}
\vspace{-0.5cm}
\caption{\label{fig:nemo}Forecasting Glorys2v4. From top to bottom: input and target observations, along with the associated ground truth partial hidden state, our model's outputs, our model variant when the initial conditions are estimated from the observations, outputs from the PKnI baseline, and from the ConvLSTM.}
\end{figure*}

\begin{table}[ht]
\caption{\label{table:mse}Quantitative results with the Glorys2v4 dataset. Mean squared error between predicted observations and ground truth for different forecast horizons $K$ (lower is better), defined as ${\frac{1}{K} \frac{1}{|\Omega|} \sum_{k=1}^{K} \sum_{x \in \Omega }\|  \H(X_k(x))-Y_k(x) \|^2}$. Note that for the ConvLSTM baseline, we do not learn interpretable states.}
\label{mse}
\vskip 0.05in
\begin{center}
\begin{small}
\begin{sc}
\begin{tabular}{lcccr}
\toprule
Model & K=5& K=10 \\
\midrule
Ours  & $0.124$ & $0.231$\\
Ours (with estimation)& $\mathbf{0.113}$ & $\mathbf{0.209}$\\
PKnI \cite{bezenac_deep_2018}    & $0.145$& $0.250$\\
ConvLSTM (\cite{convlstm})  &$0.137$ & $0.224$\\
\bottomrule
\end{tabular}
\end{sc}
\end{small}
\end{center}
\vskip -0.1in
\end{table}

\begin{table}[ht]
\caption{\label{table:similarity}Quantitative evaluation of the hidden states. Similarity between predicted hidden state and ground truth (higher is better), for different forecast horizons $K$. We use the average cosine similarity between the velocity vectors $u$ and ground truth $v$, defined as $\frac{1}{K} \sum_{k=1}^{K} \frac{1}{|\Omega|} \sum_{x \in \Omega} \frac{\left \langle u(x), v(x) \right \rangle}{\left \| u(x) \right \| \left \| v(x) \right \|}$.}
\label{similarity}
\vskip 0.05in
\begin{center}
\begin{small}
\begin{sc}
\begin{tabular}{lcccr}
\toprule
Model & K=5& K=10 \\
\midrule
Ours & $\mathbf{0.782}$ & $\mathbf{0.678}$\\
Ours (with Estimation)  & $0.77$ & $0.670$\\
PKnI \cite{bezenac_deep_2018}    & $0.448$ & $0.371$\\
ConvLSTM \cite{convlstm}    & $\times$& $\times$\\
\bottomrule
\end{tabular}
\end{sc}
\end{small}
\end{center}
\vskip -0.1in
\end{table}
\vspace*{-0.3cm}
\paragraph{Dealing with partial initial conditions.}
In order to take into account the observations made above regarding this system, especially the fact that the initial temperatures $T_0$ (in this case, since the we observe the temperatures, $ Y_0 = T_0$) and the proxy of the velocity field $\tilde{w}_0$ provided as initial input is insufficient to represent the full state, we take $g_\theta$ in equation (\ref{eq:optim_pb}) to be:
\vspace*{-0.3cm}
\begin{equation}
    g_\theta = E_\theta(Y_{-L}, \tilde{w}_0) + \begin{pmatrix}
T_0\\ 
\tilde{w}_0\\
0
\end{pmatrix}
\end{equation}

where $Y_{-L}$ corresponds to the $L$ past observations~($L=4$ in the experiments), and $E_\theta$ is an encoder neural network\footnote{In this case, $\theta$ corresponds to the parameters of $F_\theta$ and $E_\theta$, which are not shared across networks.}. Using $E_\theta$ allows us to encode available information from the observations $Y_{-L}$ which is not contained in $\breve{w}_0$ nor in $T_0$. For $E_\theta$, we use the UNet architecture~\cite{unet}. This variant accommodates our approach to model to this dataset, and shows the potential of our method to be used in settings of varying difficulty. We now compare our method against several baselines.

\subsubsection{Baselines}

\paragraph{PKnI.} This is the physics-informed deep learning model in \citet{bezenac_deep_2018}, where prior physical knowledge is integrated: it uses an advection-diffusion equation to link the velocity with the observed temperatures, and uses a neural network to estimate the velocities.
\vspace{-.3cm}
\paragraph{Convolutional LSTM.} LSTM NN which uses convolutional transitions in the inner LSTM module~\cite{convlstm}. This model can only produce observations. 

\subsubsection{Results}

We test both variants of our model. The first one is the same as in previous experiments: we take as input $(T_0, \breve{w_0})$ and consider the full state to be $X_t = (T_t, w_t)$. The second variant accommodates the fact that the latter is not the full state, and use an encoder network $E_\theta$ to produce an \textit{augmented state}. Table \ref{table:mse} shows the forecast error on the observations for different time horizons ($5$ and $10$). Note that both models variants outperform our baselines across the the different time horizons. In Table \ref{table:similarity}, we also evaluate our hidden state. For this, we calculate the cosine similarity between the hidden states associated to the proxy on the velocity vectors $\tilde{w}_t$ and the proxy itself. Interestingly, both both our methods outperform the baselines, and tend to produce vector field correlated with $\breve{w}_t$. Finally, in figure \ref{fig:nemo}, we can see that despite the high uncertainty from both the partial knowledge about the initial conditions and the varying boundary, our approach performs well.

\section{Related Work}
\label{sec:related work}

\paragraph{Data-driven Forecasting of Space-Time Dynamics.}
Forecasting space-time dynamics with machine learning methods has been a long standing endeavour. \cite{cressie2015statistics} gives a comprehensive introduction to the use of classical statistical methods to predict spatial time-series, including the use of hierarchical models. In the neural networks community, \cite{lstm} introduced the famous Long Short-Term Memory model which proved powerful in integrating temporal correlations and for which a convolutional version, more suited to spatio-temporal dependencies, was introduced by \cite{convlstm}. More recent work includes \cite{videopixel} which showed compelling results for video forecasting including on the standard Moving MNIST baseline while \cite{Ziat2017SpatioTemporalNN} used embeddings to encode the dynamics in a latent space where the forecasting is done. All the works mentionned here aimed directly to the estimation of a transition function $T$ such that $X_{t+1} = T(X_t)$ where $X$ is the studied spatial time-series which means that the dynamics aren't understood as resulting from a differential equation as we do in our approach.
\vspace{-.3cm}

\paragraph{Data-Driven Discovery of Differential Equations.}

In the past, several works have already attempted to learn differential equations from data, such as e.g. \citet{Crutchfield87equationsof}, \citet{Alvarez:2013:LLF:2554063.2554065}. More recently, \citet{rudy_data-driven_2017} uses sparse regression on a dictionary of differential terms to recover the underlying PDE. In \citet{RAISSI2017683}, they propose recovering the coefficients of the differential terms by deriving a GP kernel from a linearized form of the PDE. \citet{Long2018} carefully tailor the neural network architecture, based on the discretization of the different terms of the underlying PDE. \citet{Raissi18} develops a NN framework for learning PDEs from data. \citet{bilinear} construct a bilinear network and use an architecture similar to finite difference schemes to learn fully observed dynamical systems. In those approaches, we often see that either the form of the PDE or the variable dependency are supposed to be known and that the context is the unrealistic setting where the state is fully observed. A more hybrid example is \citet{bezenac_deep_2018} where they propose to learn a forecasting system in the partially observable case, where part of the differential equation is known, and the other is approximated using the data, which allows the network hidden state to be interpretable.

\section{Discussion}

\paragraph{Benefits of Continuous-Time.}
In the machine learning community, the forecasting problem is often seen as a learning a neural network mapping consecutive states in time. In this work, we take an alternate approach, and use the neural network to express the rate of change of the states instead. This task is intrinsically simpler for the network, and is in fact the natural way to model time varying processes. Moreover, this allows us to accommodate irregularly acquired observations, and as demonstrated by the experiments, allows interpolation between observations. From a more theoretic viewpoint, the adjoint equations derived in theorem \ref{th:theorem1} may be helpful in analyzing the behaviour of the backpropagated gradient w.r.t. the properties of the studied system.
\vspace{-.3cm}
\paragraph{Limitations.}
However, there are still many aspects to explore. The fact that we are using explicit discretization should be limiting w.r.t. the class of equations we can learn as stiff equations necessitate the use of implicit methods and this can be worked around by the adjoint method we presented. We have also restricted ourselves to a linear $\H$ and it would be interesting to see how our algorithms work for operators with a more complicated structure. Finally, we have restricted ourselves to the stationary hypothesis while, as we can see through the Glorys2v4 example, real-world processes, when looked at from a local point of view\footnote{Meaning that not all exterior forces are factored into the model.}, aren't. These are interesting directions for future work.
\vspace{-.3cm}
\paragraph{Hidden State Discovery.}
By feeding the initial condition to the neural network, and training the network to regress only to the observations, it was not expected that the neural network would forecast the hidden state in a way that closely mimics the true state of the underlying dynamical system. Indeed, the neural network must predict a hidden state that contains the information of the dynamical system's state in order to correctly forecast the observations for multiple time steps, but the way the network structures this information is not constrained by the loss functional. We believe that these results are due to the fact that is easier for the network to use the same coding scheme as in the initial condition, instead of creating a disjoint code of its own for the following time steps. We see this empirical result as a very important one as it implies that it is possible to learn very complex dynamics with only partial information, without necessarily incorporating prior knowledge on the dynamics of the state. Along with the results obtained for the very challenging Glorys2v4 dataset, we are convinced this constitutes an important step towards applying learning to real-world physical processes. Obviously, the interaction of this phenomenon with the integration of physical priors into the algorithm, for example by adding explicit differential operators into $F$, is a very interesting question.

\vspace{-.3cm}
\section{Conclusion}

We have introduced a general data-driven framework to predict the evolution of space-time processes, when the system is highly complex and nonlinear and the state is not fully observed. Assuming the underlying system follows a time-dependant differential equation, we estimate the unknown evolution term with a neural network. We argue that this is in fact a natural way to model continuous-time systems.
Viewing its parameters as control variables, we propose a learning algorithm for the neural network, making use of results from continuous-time optimal control theory. Experiments performed on two simulated datasets from fluid dynamics and on data from a sophisticated data simulator used in climate modeling show that the proposed method not only is able to produce high quality forecasts at different horizons, but also learns with a good accuracy the underlying state space dynamics. This may open the way for new methods for integrating prior physical knowledge, \textit{e.g.} by imposing constraints directly on the modeled evolution term.

\bibliography{biblio}
\bibliographystyle{icml2019}

\clearpage
\newpage
\newpage
\appendix

\section{Equations}

In this section, we succinctly describe the equations used in our experiments.

\subsection{The Shallow Water Equations\label{shalw}}

The shallow-water model can be written as:
\begin{eqnarray}
\dfrac{\partial u}{\partial t} & = &
+ (f + \zeta ).v  - \partial_x (\frac{u^2+v^2}{2} + g^*.h) + \nonumber\\ &&\frac{\tau_x}{\rho_0(H+h)} - \gamma . u + \nu\Delta u \nonumber \\
\dfrac{\partial v}{\partial t} & = &
-  (f + \zeta ).u  - \partial_y (\frac{u^2+v^2}{2} + g^*.h) + \nonumber\\
                  &&\frac{\tau_y}{\rho_0(H+h)} -
                                       \gamma . v + \nu\Delta v \\
\dfrac{\partial h}{\partial t} & = &
- \partial_x(u(H+h)) - \partial_y(v(H+h)) \nonumber
\end{eqnarray}

where:
\begin{itemize}
\item[--] $u$, $v$, $h$ are state variables, standing for velocity and mixed layer depth anomaly)
\item[--] $\zeta$  is the vorticity.
\item[--] $g^*=$\SI{0.02}{\meter \per \second \squared} is the reduced gravity
\item[--] $H=500m$ is the mean mixed-layer depth.
\item[--] $\rho_0$ is the density of the water set to $1000 mg/m^3$
\item[--] $\gamma$ is the dissipation coefficient set to $2\cdot10^{-7}s^{-1}$
\item[--] $\nu$ is the diffusion coefficient set to $0.72 m^2/s$
\item[--] $\tau_x$ is the zonal wind forcing defined in Eq.~\ref{eq:wind}
\end{itemize}
The zonal wind forcing is defined as:
\begin{equation*}
\tau_x(y) = \tau_0 \sin(2\pi (y-y_c)/L_y
\label{eq:wind}
\end{equation*}
where:
\begin{enumerate}

\item[--] $\tau_0$ is the maximum intensity of the wind stress(in the standard case $0.15 m.s^{-2}$).
\item[--]$y$ is the latitude coordinate
\item[--] $y_c$ is the center $y$ coordinate of the domain
\item[--] $L_y$ is the length of the domain ($L_y = 1600 km$ in our case).
\end{enumerate}

Here, the state is composed of the velocity vector and the mixed layer depth:
\[
X = 
\begin{pmatrix}
u\\v\\h
\end{pmatrix} \; \text{and} \; \; \H(X) = h
\]

For our simulations, the spatial differential operators have been discretized using finite differences on a Arakawa C-grid.

\subsection{The Euler Equations\label{section:euler}}

\begin{equation}
\label{eq:euler}
      \begin{aligned}
        \dfrac{\partial u}{\partial t} + (u\cdot\nabla)u = -\dfrac{\nabla p}{\rho} + g\\
        \dfrac{\partial \rho}{\partial t} + (u\cdot\nabla)\rho = 0\\
        \nabla\cdot u = 0
      \end{aligned}
\end{equation}
where $\nabla\cdot$ is the divergence operator, $u$ corresponds to the flow velocity vector, $p$ to the pressure, and $\rho$ to the density.

The Euler equations are not of the form \eqref{eq:state} as we still have the pressure variable $p$ as well as the null divergence constraint. However, the Helmholz-Leray decomposition result states that for any vector field $a$, there exists $b$ and $c$ such that~:
\[
a = \nabla b + c \]
and
\[
\nabla\cdot c = 0
\]

Moreover, this pair is unique up to an additive constant for $b$. Thus, we can define a linear operator $\mathbb{P}$ by~:
\[
\mathbb{P}(a) = c
\]
This operator is a continuous linear projector which is the identity for divergence-free vector fields and vanishes for those deriving from a potential.

By taking a solution of \eqref{eq:euler} and applying $\mathbb{P}$ on the first equation, we have, as $u$ is divergence free from the third equation and as $g$ derives from a potential~:
\[
\dfrac{\partial u}{\partial t} = - \mathbb{P}[(u\cdot\nabla)u]
\]
where permuting derivation and $\mathbb{P}$ is justified by the continuity of the operator\footnote{One can use a finite difference approximation to show it for example.}.

Thus, if $u$ is solution to \eqref{eq:euler}, it is also a solution of~:
\begin{equation*}
      \begin{aligned}
        \dfrac{\partial u}{\partial t} = - \mathbb{P}[(u\cdot\nabla)u] \\
        \dfrac{\partial \rho}{\partial t} = -(u\cdot\nabla) \rho\\
      \end{aligned}
\end{equation*}
which is of the form of \eqref{eq:state}.

Conversely, the solution of the above system is such that~:
\[
u_t = \int \dfrac{\partial u}{\partial t} = \int - \mathbb{P}[(u\cdot\nabla)u]
\]
which gives, by exchanging $\mathbb{P}$ and the integral\footnote{To prove this, we can take a sum approximation to the integral and use again the linearity then the continuity of $\mathbb{P}$.}~:
\[
u_t = \mathbb{P}\left[ \int - (u\cdot\nabla)u \right]
\]
so that $u$ is automatically of null divergence by definition of $\mathbb{P}$. The two systems are thus equivalent.

In conclusion, we have:
\[
X = 
\begin{pmatrix}
u\\ \rho
\end{pmatrix}, \, \text{and} \; \: \H(X) = \rho
\]

Moreover, $u$ is generally a two or three-dimensional spatial field while $\rho$ is a scalar field.

\subsection{Glorys2v4\label{section:nemo}}
The Glorys2v4 product is a reanalysis of the global Ocean (and the Sea Ice, not considered in this work). The numerical ocean model is NEMOv3.1~\cite{madec} constrained by partial real observations of Temperature, Salinity and Sea Level. Oceanic output variables of this product are daily means of Temperature, Salinity, Currents, Sea Surface Height at a resolution of 1/4 degree horizontal resolution.

The NEMO model describes the ocean by the primitive equations (Navier-Stokes equations together with an equation of states). 

Let $(\mathbf{i}, \mathbf{j}, \mathbf{k})$ the 3D basis vectors, $U$ the vector velocity, ${\bf U} = {\bf U}_h + w \mathbf{k}$ (the subscript $h$ denotes the local horizontal vector, \textit{i.e.} over the $(\mathbf{i}, \mathbf{j})$ plane), $T$ the potential temperature, $S$ the salinity, $\rho$ the \textit{in situ} density. The vector invariant form of the primitive equations in the $(\mathbf{i}, \mathbf{j}, \mathbf{k})$ vector system provides the following six equations (namely the momentum balance, the hydrostatic equilibrium, the incompressibility equation, the heat and salt conservation equations and an equation of state):

\begin{gather*} 
	\frac{\partial {\bf U}_h}{\partial t}  = - \bigg[ ({\bf U} . \nabla) {\bf U} \bigg]_h - f \mathbf{k} \times {\bf U}_h - \frac{1}{\rho_0} \nabla_h p + D^{\bf U} + F^{\bf U} \\
	\frac{\partial p}{\partial z} = - \rho g \\
	\nabla . {\bf U} = 0 \\
	\frac{\partial T}{\partial t} = - \nabla . (T {\bf U}) + D^T + F^T \\
	\frac{\partial S}{\partial t} = - \nabla . (S {\bf U}) + D^S + F^S \\
	\rho = \rho(T, S, p)
    \label{eq:nemo_state}
\end{gather*} 

where $\rho$ is the \textit{in situ} density given by the equation of the state \ref{eq:nemo_state}, $\rho_0$ is a reference density, $p$ the pressure, $f = 2 \Omega . \mathbf{k}$ is the Coriolis acceleration. $D^U$, $D^T$ and $D^S$ are the parameterizations of small-scale physics for momentum, temperature and salinity, and $F^U$, $F^T$ and $F^S$ surface forcing terms.

As in subsection \ref{section:euler}, the divergence-free constraint over can be enforced through the Leray operator. Moreover, $\rho$ is a function of other state variables so that the state can be written as:
\[
X =
\begin{pmatrix}
U\\ p \\S \\T
\end{pmatrix} \; \text{and} \; \; \H(X) = \overline{T}.
\]
where $\overline{T}$ is the daily mean temperature derived from the instantaneous potential temperature T in the model.
\clearpage
\newpage

\section{Proof of Theorem \ref{th:theorem1}\label{proof}}

We start by differentiating $\La$. In what follows, all considered functions are supposed to be twice continuously differentiable in all variables and we will use the notation $\partial_u F(u_0)$ to designate the differential of $F$ with respect to $u$ \textit{i.e.} the unique linear operator such that:
\[
F(u_0+\delta u) = F(u_0) + \partial_u F(u_0)\delta u + o(\delta u)
\]
By hypothesis, we consider this operator to be always continuous in our case.

Straightforward calculus gives us:
\[
\dfrac{\partial\J(X^\theta_t)}{\partial\theta} = \int_0^T 2\left\langle \partial_X\H(X^\theta_t)\cdot\partial_\theta X^\theta_t, \H(X^\theta_t) - Y_t \right\rangle \mathrm{dt}
\]

Let us fix $\theta$ and a variation $\delta \theta$. Then, we have, by definition:
\[
X^{\theta+\delta\theta} = X^\theta_t + \partial_\theta X^\theta_t \cdot\delta\theta + o(\delta\theta)
\]
and, for any $X$ and any $\delta X$:
\[
F_\theta(X+\delta X) = F(X) + \partial_X F_\theta(X) \cdot \delta X + o(\delta X)
\]
and:
\[
F_{\theta+\delta\theta}(X) = F_\theta(X) + \partial_\theta F_\theta(X)\cdot\delta\theta + o(\delta\theta)
\]
so that:
\[
F_{\theta+\delta\theta}(X^{\theta+\delta\theta}_t) = F_\theta(X^{\theta+\delta\theta}_t) + \partial_\theta F_\theta(X^{\theta+\delta\theta}_t)\cdot\delta\theta + o(\delta\theta)
\]
Then, because $F$ is twice continuously differentiable:
\begin{equation*}
\begin{split}
\partial_\theta F_\theta(X^{\theta+\delta\theta}_t) &= \partial_\theta F_\theta\left(X^\theta_t+\partial_\theta X^{\theta}_t\cdot\delta\theta + o(\delta\theta)\right) \\&= \partial_\theta F_\theta(X^\theta_t) + \partial_X \partial_\theta F_\theta(X^\theta_t)\cdot\partial_\theta X^{\theta}_t\cdot\delta\theta \\&+ o(\delta\theta)
\end{split}
\end{equation*}
and:
\begin{equation*}
\begin{split}
F_\theta(X^{\theta+\delta\theta}_t) &= F_\theta\left(X^{\theta}_t+\partial_\theta X^{\theta}_t\cdot\delta\theta + o(\delta\theta)\right) \\&= F_\theta(X^{\theta}_t)+\partial_X F_\theta(X^\theta_t)\cdot\partial_\theta X^{\theta}_t\cdot\delta\theta + o(\delta\theta)
\end{split}
\end{equation*}
Moreover, as all differential operators below are continuous by hypothesis, we have that:
\[
\|(\partial_X \partial_\theta F_\theta(X^\theta_t)\cdot\partial_\theta X^{\theta}_t\cdot\delta\theta)\cdot\delta\theta\| \leq\|\partial_X \partial_\theta F_\theta(X^\theta_t)\|\  \|\partial_\theta X^{\theta}_t\|\ \|\delta\theta\|^2
\]
so that:
\begin{equation*}
\begin{split}
F_{\theta+\delta\theta}(X^{\theta+\delta\theta}_t) &\\= F_\theta(X^{\theta}_t)&+\left(\partial_X F_\theta(X^\theta_t)\cdot\partial_\theta X^{\theta}_t + \partial_\theta F_\theta(X^\theta_t)\right)\cdot\delta\theta + o(\delta\theta)
\end{split}
\end{equation*}

We now have all elements to conclude calculating the derivative of $\La$, with some more easy calculus:
\begin{equation*}
\begin{split}
\dfrac{\partial \La}{\partial\theta} = \int_0^T &\left(2\left\langle \partial_X\H(X^\theta_t)\cdot\partial_\theta X^\theta_t, \H(X^\theta_t) - Y_t \right\rangle\right. + \\&\left.\left\langle \lambda_t, \partial_\theta\partial_t X^\theta_t-\partial_X F_\theta(X^\theta_t)\cdot\partial_\theta X^{\theta}_t - \partial_\theta F_\theta(X^\theta_t) \right\rangle \right)\mathrm{dt} \\&+ \left\langle \mu, \partial_\theta X^\theta_0-\partial_\theta g_\theta \right\rangle
\end{split}
\end{equation*}

By the Schwarz theorem, as $X$ is twice continuously differentiable, we have that $\partial_\theta\partial_t X^\theta_t = \partial_t\partial_\theta X^\theta_t$. Integrating by parts, we get:
\begin{equation*}
\begin{split}
\int_0^T \left\langle \lambda_t, \partial_\theta\partial_t X^\theta_t \right\rangle\mathrm{dt} = &\left\langle \lambda_T,\partial_\theta X^\theta_T \right\rangle - \left\langle \lambda_0,\partial_\theta X^\theta_0 \right\rangle  \\&- \int_0^T \left\langle \partial_t\lambda_t, \partial_\theta X^\theta_t \right\rangle \mathrm{dt}
\end{split}
\end{equation*}

Putting all this together and arranging it, we get:
\begin{equation*}
\begin{split}
\dfrac{\partial \La}{\partial\theta} &=  \int_0^T \left\langle \partial_\theta X^\theta_t, 2\partial_X\H(X^\theta_t)^\star\left(\H(X^\theta_t) - Y_t\right)\right. \\& \left.- \partial_t\lambda_t - \partial_X F_\theta(X^\theta_t)^\star\lambda_t \right\rangle\mathrm{dt} \\&- \int_0^T \left\langle \lambda_t, \partial_\theta F_\theta(X^\theta_t) \right\rangle \mathrm{dt} + \left\langle \lambda_T,\partial_\theta X^\theta_T\right\rangle +  \left\langle\mu-\lambda_0,\partial_\theta X^\theta_0\right\rangle \\&- \left\langle \mu,\partial_\theta g_\theta \right\rangle
\end{split}
\end{equation*}

We can now define:
\[
A_t = -(\partial_X F_\theta(X_t^\theta))^\star
\]
and
\[
B_t = 2(\partial_X\H(X_t^\theta))^\star(\H(X_t^\theta)-Y_t)
\]
and, recalling that $\lambda$ can be freely chosen, impose that $\lambda$ is solution of:
\[
\partial_t\lambda_t = A_t\lambda_t + B_t
\]
with final condition $\lambda_T=0$.
We also choose $\mu = \lambda_0$ so that, finally, we have:
\[
\dfrac{\partial \La}{\partial\theta} =  - \int_0^T \left\langle \lambda_t, \partial_\theta F_\theta(X^\theta_t) \right\rangle \mathrm{dt} - \left\langle \lambda_0,\partial_\theta g_\theta \right\rangle
\]
which concludes the proof.\qed 

\clearpage
\newpage
\onecolumn
\section{Additional Forecasts}
\begin{figure*}[h!]
\captionsetup{width=0.65\textwidth}
\begin{center}
\includegraphics[width=0.55\textwidth]{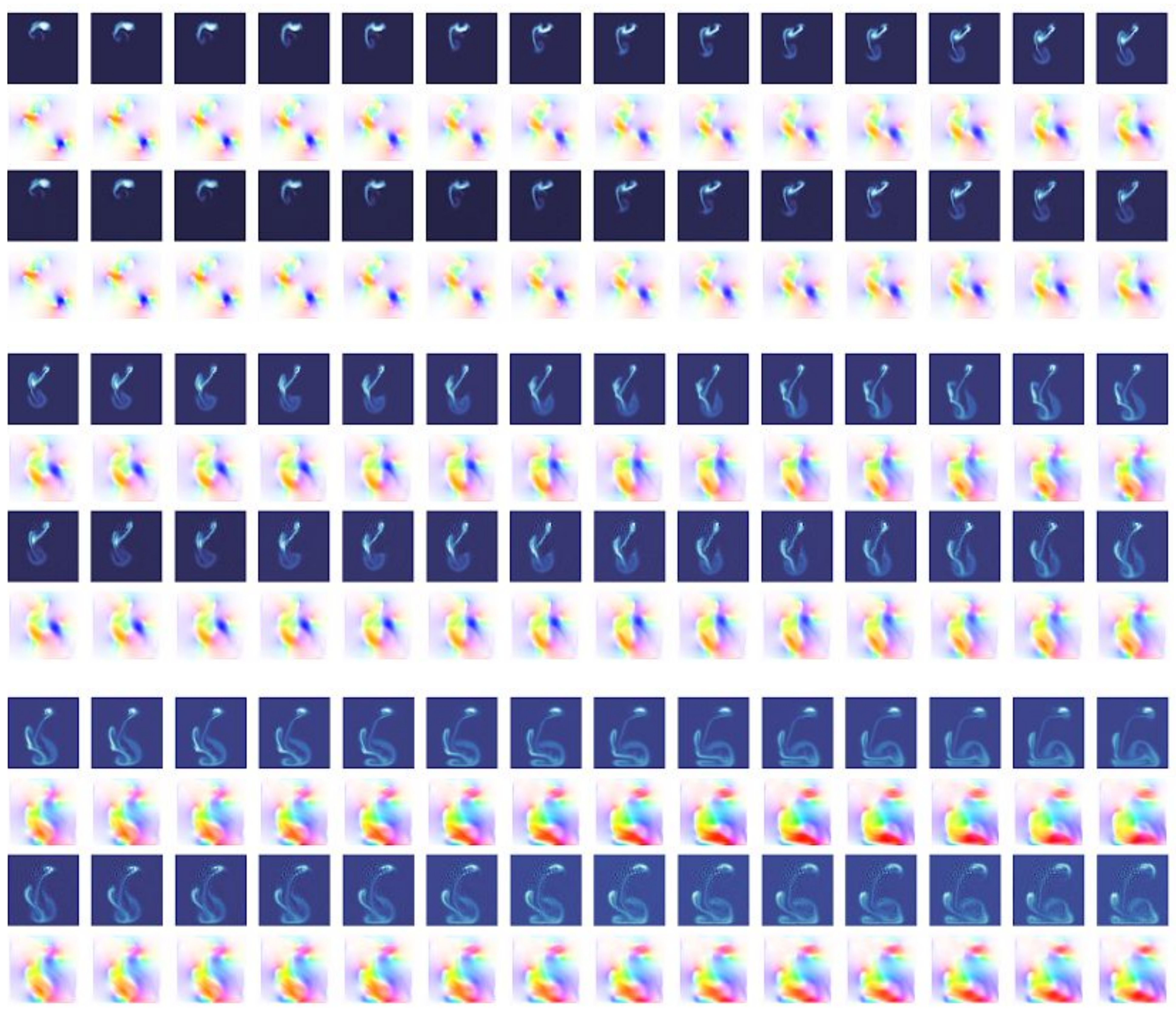}
\caption{Forecasting the Euler equations, starting from a given initial condition (not shown here). We forecast 42 time-steps ahead (rows $0,1(\!\!\!\!\mod4)$) and compare results with the ground truth simulation (rows $2,3(\!\!\!\!\mod4)$).}
\end{center}
\end{figure*}

\begin{figure*}[h!]
\captionsetup{width=0.65\textwidth}
\begin{center}\includegraphics[width=0.55\textwidth]{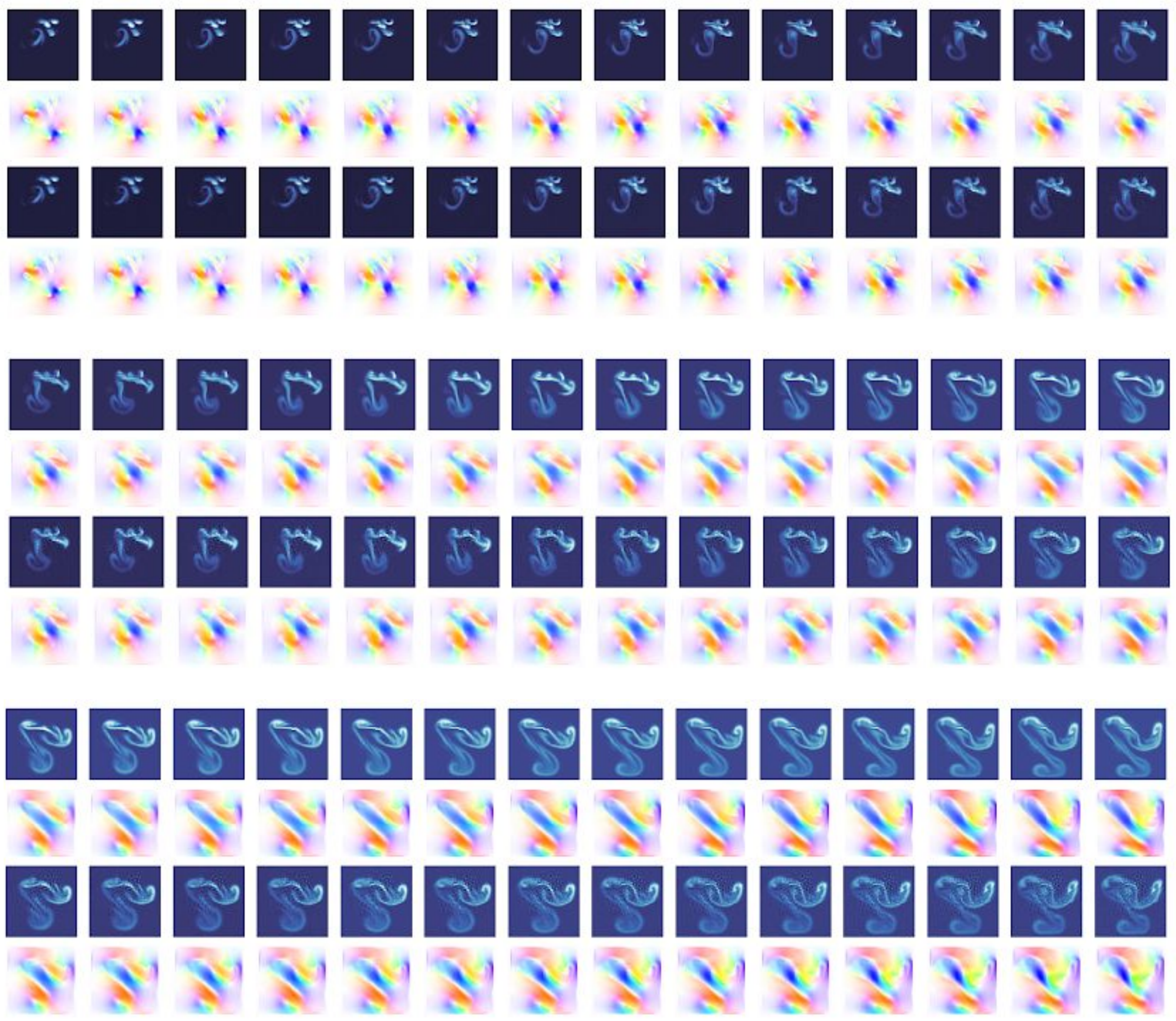}
\caption{Forecasting the Euler equations, starting from a given initial condition (not shown here). We forecast 42 time-steps ahead (rows $0,1(\!\!\!\!\mod4)$) and compare results with the ground truth simulation (rows $2,3(\!\!\!\!\mod4)$).}
\end{center}
\end{figure*}

\begin{figure*}
\captionsetup{width=.65\linewidth}
\begin{center}
\includegraphics[width=0.55\textwidth]{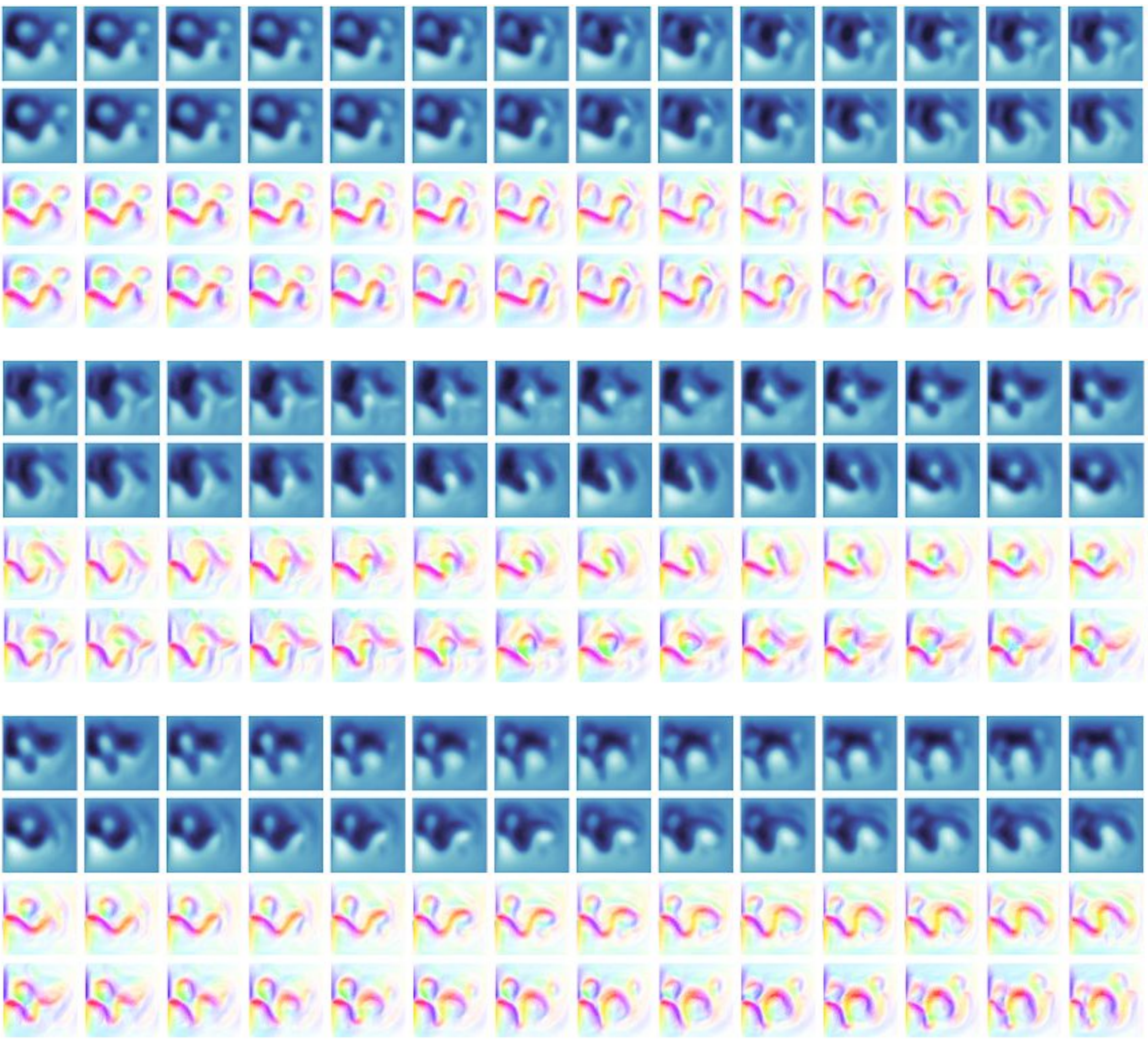}
\end{center}
\caption{Forecasting the shallow water equations, starting from a given initial condition (not shown here). We forecast 42 time-steps ahead (rows $0,1(\!\!\!\!\mod4)$) and compare results with the ground truth simulation (rows $2,3(\!\!\!\!\mod4)$).}
\end{figure*}

\begin{figure*}
\captionsetup{width=.65\linewidth}
\begin{center}
\includegraphics[width=0.55\textwidth]{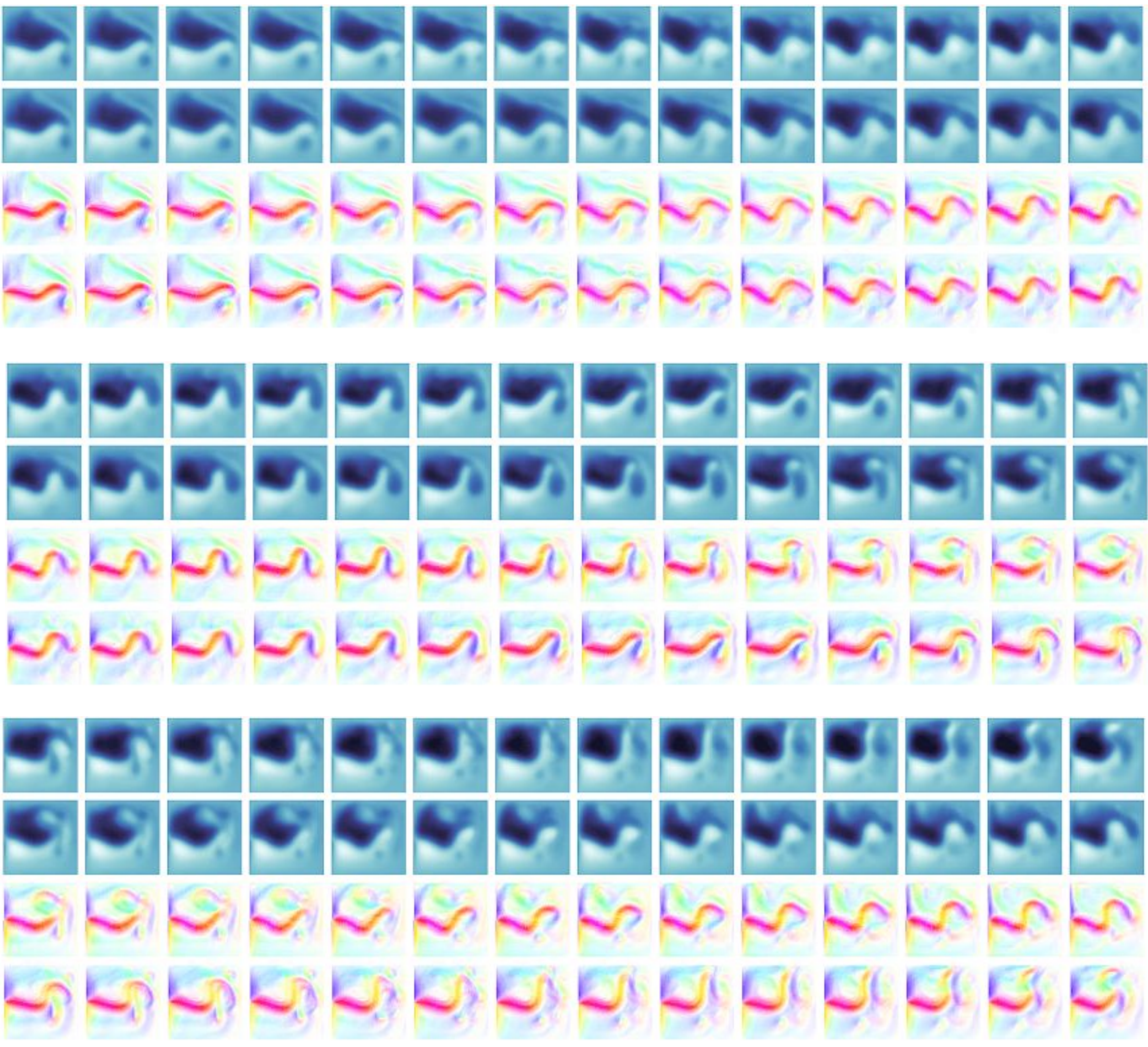}
\end{center}
\caption{Forecasting the shallow water equations, starting from a given initial condition (not shown here). We forecast 42 time-steps ahead (rows $0,1(\!\!\!\!\mod4)$) and compare results with the ground truth simulation (rows $2,3(\!\!\!\!\mod4)$).}
\end{figure*}

\begin{figure*}
\captionsetup{width=.55\linewidth}
\begin{center}
\includegraphics[width=0.5\textwidth]{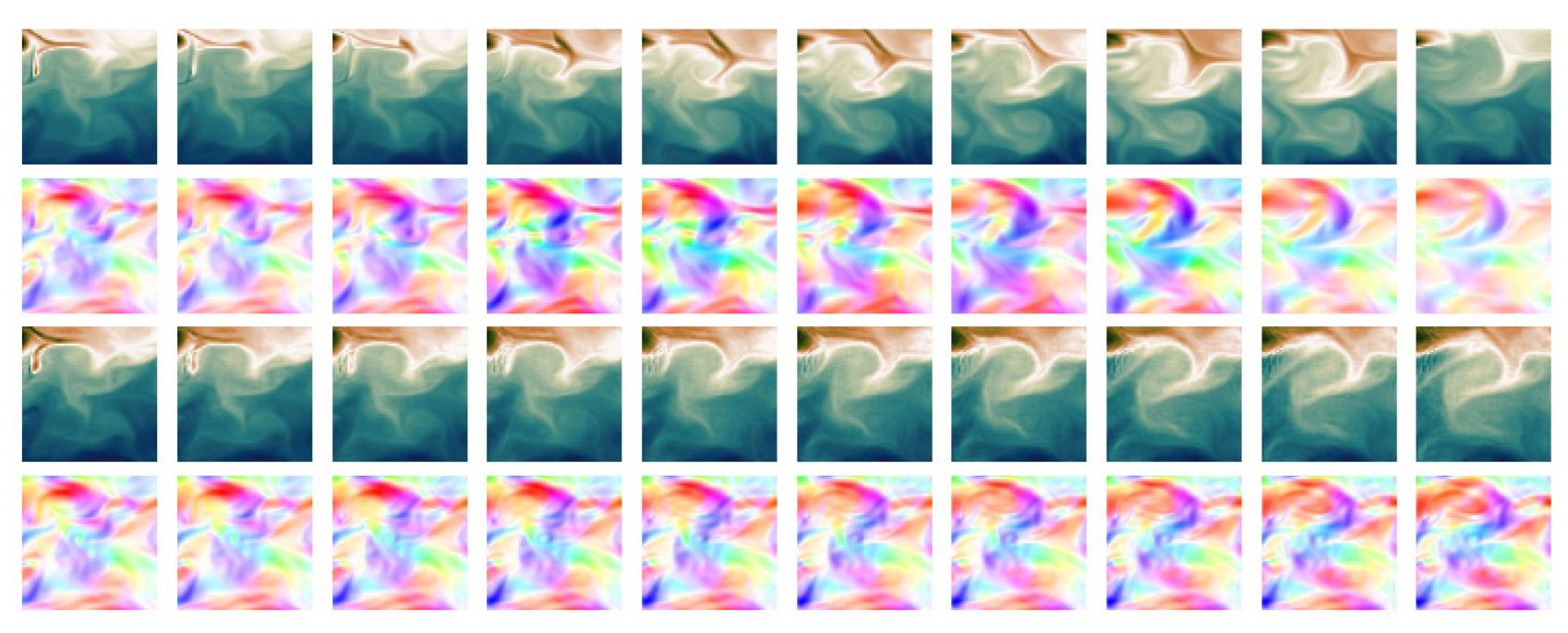}
\end{center}
\caption{Forecasting Glorys2v4 $10$ time-steps ahead, starting from a given initial condition (not shown here). Top two rows: ground truth, bottom two rows: model forecasts.}
\end{figure*}
\vspace{-1cm}

\begin{figure*}
\captionsetup{width=.55\linewidth}
\begin{center}
\includegraphics[width=0.5\textwidth]{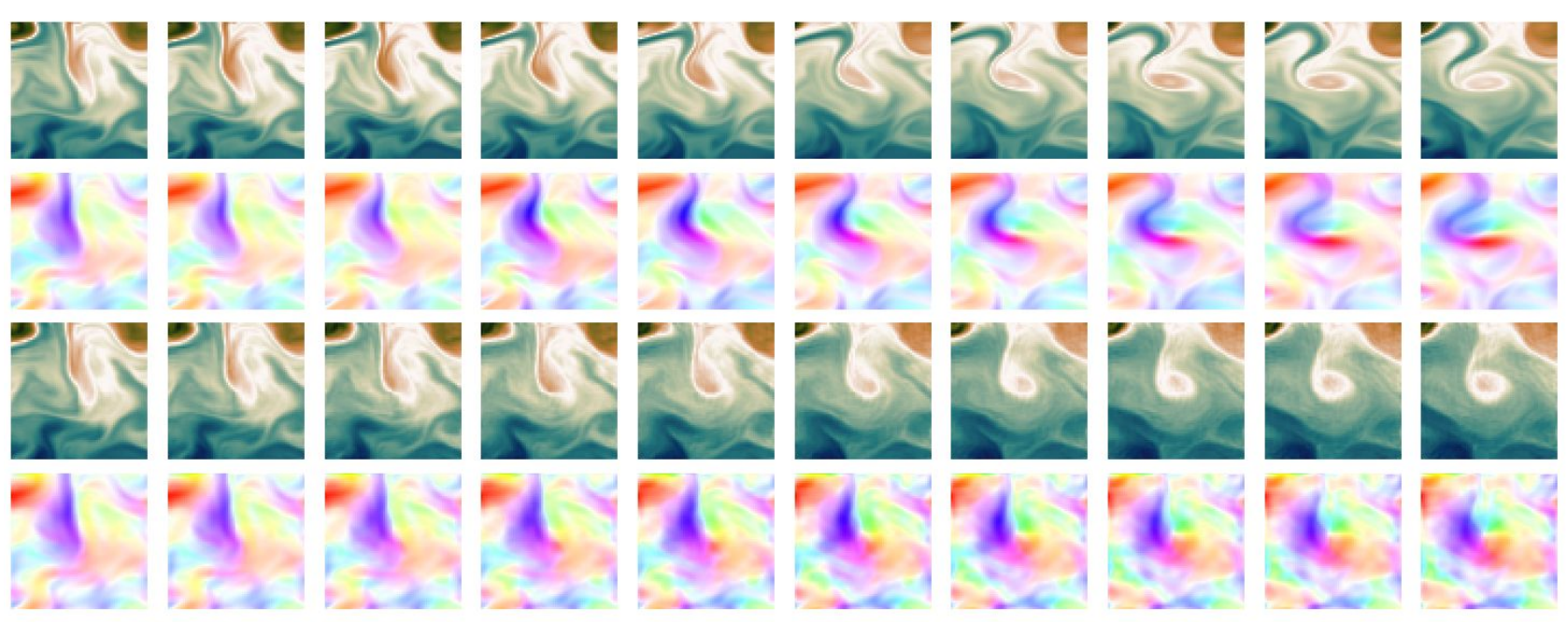}
\end{center}
\caption{Forecasting Glorys2v4 $10$ time-steps ahead, starting from a given initial condition (not shown here). Top two rows: ground truth, bottom two rows: model forecasts.}
\end{figure*}
\vspace{-1cm}

\begin{figure*}
\captionsetup{width=.55\linewidth}
\begin{center}
\includegraphics[width=0.5\textwidth]{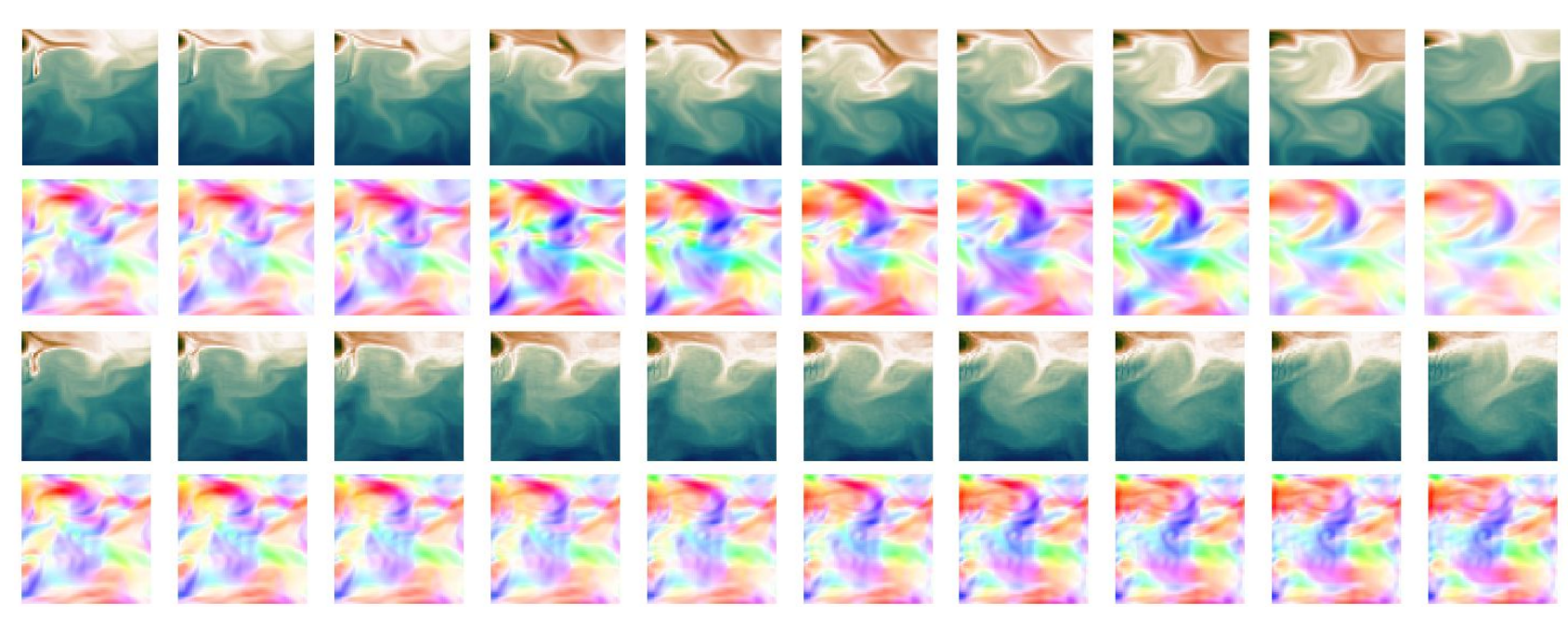}\\
\end{center}
\caption{Forecasting Glorys2v4 $10$ time-steps ahead with estimation step, starting from a given initial condition (not shown here). Top two rows: ground truth, bottom two rows: model forecasts.}
\end{figure*}
\vspace{-1cm}

\begin{figure*}
\captionsetup{width=.55\linewidth}
\begin{center}\includegraphics[width=0.5\textwidth]{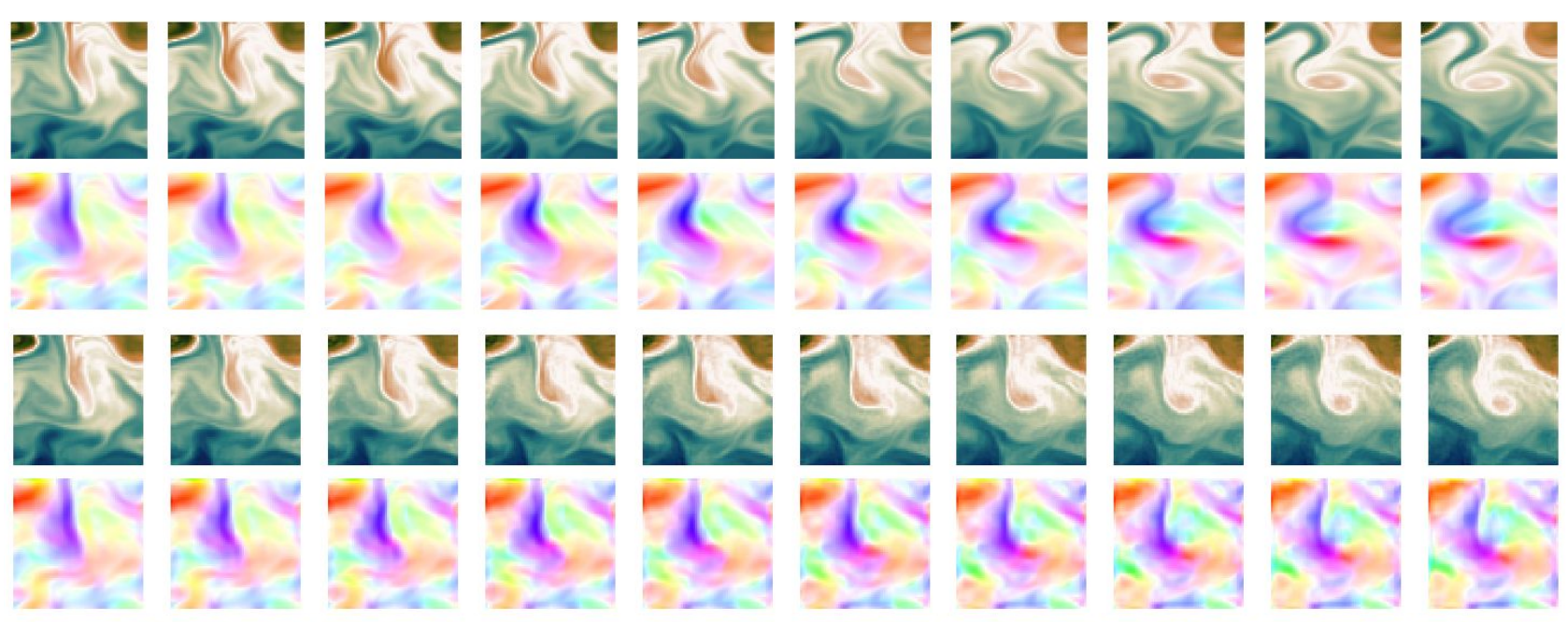}
\end{center}
\caption{Forecasting Glorys2v4 $10$ time-steps ahead with estimation step, starting from a given initial condition (not shown here). Top two rows: ground truth, bottom two rows: model forecasts.}
\end{figure*}

\end{document}